\documentclass[traditabstract]{aa}  
\usepackage{graphicx}
\usepackage{natbib}
\begin{document}

   \title{A long-period massive planet around HD106515A
          \thanks{Based on observations made with the Italian Telescopio 
                  Nazionale Galileo (TNG) operated on the island of La Palma 
                  by the Fundacion Galileo Galilei of the INAF 
                  (Istituto Nazionale di Astrofisica) at the Spanish 
                  Observatorio del Roque de los Muchachos of the Instituto 
                  de Astrofisica de Canarias.}}

\titlerunning{A long-period massive planet around HD106515A}

   \author{S. Desidera
           \inst{1},
           R. Gratton
           \inst{1},
           E. Carolo
           \inst{1,2},
           A.F. Martinez Fiorenzano
           \inst{3},
           M. Endl
           \inst{4},
           D. Mesa
           \inst{1},
           M. Cecconi
           \inst{3},
           R. Claudi
           \inst{1},
           R. Cosentino
           \inst{3,5},
           S. Scuderi
           \inst{5},
           A. Sozzetti
           \inst{6},
           A. Zurlo
           \inst{1,7}
           }

   \authorrunning{S. Desidera et al.}

   \offprints{S. Desidera,  \\
              \email{silvano.desidera@oapd.inaf.it} }

   \institute{INAF -- Osservatorio Astronomico di Padova,  
              Vicolo dell' Osservatorio 5, I-35122, Padova, Italy
             \and 
             Dipartimento di Astronomia -- Universit\'a di Padova, Vicolo
             dell'Osservatorio 2, Padova, Italy 
             \and
             INAF -- Centro Galileo Galiei, Calle Alvarez de Abreu 70, 38700 
             Santa Cruz de La Palma (TF), Spain 
             \and
             McDonald Observatory, The University of Texas at Austin, Austin, 
             TX 78712, USA
            \and
             INAF -- Osservatorio Astrofisico di Catania, 
             Via S.Sofia 78, Catania, Italy
             \and
             INAF -- Osservatorio Astrofisico di Torino, via Osservatorio 20, 10025 Pino Torinese, Italy
             \and
             Aix Marseille Universit\'e, CNRS, LAM (Laboratoire d'Astrophysique de Marseille) 
             UMR 7326, 13388, Marseille, France}

\date{Received  / Accepted }

\abstract{We have performed RV monitoring of the components of the binary 
system HD 106515 over about 11 years using the high resolution spectrograph 
SARG at TNG.
The primary shows long-period radial velocity variations that indicate
the presence of a low mass companion whose projected mass is in the
planetary regime ($m \sin i = 9.33~M_{J}$).  The 9.8 years orbit results 
quite eccentric ($e=0.57$), as typical for massive giant planets.
Our results confirm the preliminary announcement of the planet included in 
Mayor et al.~(2011).
The secondary instead does not show significant RV variations.
The two components do not differ significantly in chemical composition,
as found for other pairs for which one component hosts giant planets.
Adaptive optics images obtained with AdOpt@TNG do not reveal additional
stellar companions.
From the analysis of the relative astrometry of the components of the wide pair 
we put an upper limit on the mass of the newly detected companion of about
$0.25~M_{\odot}$. State of art or near future instrumentation can provide
true mass determination, thanks to the availability of the wide companion 
HD106515B as reference. Therefore, HD106515Ab will allow deeper insight in the
transition region between planets and brown dwarfs.}

   \keywords{(Stars:) individual: HD 106515 - Planetary systems - (Stars:) binaries: visual  - 
              Techniques: spectroscopic - (Stars:) brown dwarfs - Techniques: high angular resolution}

   \maketitle

\section{Introduction}
\label{s:intro}

The upper mass limit of planetary objects is currently widely debated in the
scientific community.
On one hand, an operational definition can be based 
on the minimum mass for deuterium burning (about $13~M_{J}$)
as a dividing line between planets and brown dwarfs
\citep{2001RvMP...73..719B}, as defined by IAU \citep{2012IAUTB..28..138B}, 
or to different fixed threshold values 
\citep[e.g. $24-25~M_{J}$:][]{2006ApJ...646..505B,2011A&A...532A..79S}.
On the other hand, a definition based on formation mechanisms
is more difficult to obtain. In fact, observational data
are incomplete and the detection techniques in most cases
can provide only indirect indication on the formation of a detected substellar
objects. Furthermore, only minimum masses are known for most of the objects detected by 
RV surveys or, in the case of objects 
detected through direct imaging,  true masses are uncertain because of the 
large sensitivity of luminosity on age and intrinsic uncertainties
of theoretical models especially at young ages
\citep{2003A&A...402..701B}.

In spite of these difficulties there is growing evidence for
a significant overlap in mass between objects formed like stars do
and objects formed in a protoplanetary disks.
Several objects of planetary mass have been detected as free floating objects
in star clusters, star forming regions or in the field using imaging 
\citep[e.g.][]{2000Sci...290..103Z,2012ApJ...748...74L,2012arXiv1207.1449S}
and microlensing
\citep{2011Natur.473..349S}
or as very wide companions of stars \citep[e.g.][]{2005A&A...438L..29C}.
Some of these objects might be formed in planetary systems
and then pushed at very wide separation or ejected from the system
because of dynamical interactions with other planets 
\citep[Jumping Jupiters scenario,][]{2002Icar..156..570M}.
However, it seems unlikely this is the only mechanism producing free floating objects
below deuterium burning mass \citep{2011ApJ...743..148B}. 
Instead, the minimum mass for core collapse was found to be of a few $M_{J}$ 
and is then likely that objects of planetary mass formed star-like 
outside planetary disks \citep{2007prpl.conf..459W}.
The statistics of low mass brown dwarfs appear to be compatible with Jeans mass
fragmentation of an interstellar molecular cloud \citep{2009A&A...493.1149Z}.

On the other hand, there are objects with masses from $13$ to about 
$25~M_{J}$ that are found in systems with other lower mass planets, such as
HD168443 and HAT-P13. In the latter case the planetary nature of the lower mass companion
is confirmed by the occurrence of transits \citep{2009ApJ...707..446B}. 
In other cases such as HD~38519 and HD~202206 
a debris disk is present in the system
beside a massive planet and a lower mass planet \citep{2010ApJ...717.1123M}. 
These facts support the formation
of these objects in a protoplanetary disks.
Planet formation models also predict the presence of very massive
planets, up to about $38~M_{J}$, in exceptional cases of long-lived, massive 
and metal-rich disks \citep{2009A&A...501.1139M}.
The rising mass function below about $20-30~M_{J}$ \citep{2006ApJ...640.1051G}
is another indication that a different formation mechanism starts to be 
present above deuterium burning mass.

This picture is further complicated by evidences that the
statistical properties of planetary mass companions with (projected) masses
between  $4$ to $15~M_{J}$ are different from those of lower mass planets
\citep{2007A&A...464..779R}.
It is then possible that either a different formation mechanism is
in action or that the evolution of massive planets in a protoplanetary disk
is different depending on planetary mass.

The statistic of substellar objects in the mass range between $10$ to 
$30~M_{J}$ is still limited, given the intrinsic rarity of these objects 
\citep{2010A&A...509A.103S,2012A&A...538A.113D}.
The discovery of additional candidates is then welcome, especially when
the true mass of the companion can be determined or significantly constrained through
astrometry \citep{2011A&A...525A..95S,2011A&A...527A.140R}.
We present here the confirmation of a high-mass planet candidate
orbiting the star HD106515A and clues on its mass from astrometry.
The planet was first included in the compilation by 
\cite{2011arXiv1109.2497M} but only orbital period, RV semiamplitude,
eccentricity, and corresponding minimum masses and semimajor axis
are listed, postponing a more detailed analysis to a forthcoming paper.
HD 106515A is part of a wide binary system, with the companion
HD 106515B at a projected separation of about 250 AU.
Both components were observed as part of the RV survey
looking for planets around the components of moderately wide
binaries performed using SARG at TNG.

\section{Observations and data reduction}
\label{s:obs}

Observations were performed at the Italian Telescopio
Nazionale Galieleo (TNG) using the high resolution spectrograph
SARG \citep{2001ExA....12..107G}.
All but one spectra, used as template in the RV determination
process, were acquired with the iodine cell inserted in the optical
path. The observing procedure and the instrument set-up
are the standard ones for SARG planet search survey.
We refer to \cite{2011A&A...533A..90D} for further details.
In the case of HD106515, the chance of contamination of the spectra
is negligible even in the worst observing conditions, 
thanks to the 6.8 arcsec separation on the sky between the 
components.

The acquisition of data on HD106515 was stopped after May 2009
because of lack of time allocation and was recently restarted after
the publication of \cite{2011arXiv1109.2497M}, that listed
a planet around the primary component of this system.
The 900s integration yields typical S/N ratios of 100.
for \object{HD106515A} and 90 for \object{HD106515B}.

Differential radial velocities were obtained using the AUSTRAL code
\citep{2000A&A...362..585E}, achieving internal errors of about 3 m/s.
Data taken in 2011-2012 have a somewhat larger uncertainty, because of 
significant asymmetries of the spectrograph instrument profile.
Standard RV analysis performed using a central Gaussian with 
two or four Gaussian satellites show RV higher by few tens of m/s 
with respect to previous data. However, when analyzing the data exploiting the
Maximum Entropy algorithm such discrepancy vanishes. 
In any case, while the internal errors are similar to the older data, we
can not exclude the occurrence of systematics at 10-15 m/s level for these
recent data. The RV time series of HD106515A and B are reported 
in Table~\ref{t:rva} and \ref{t:rvb}.

HD 106515 was also observed on 21 June 2007 using AdOpt@TNG 
\citep{2006SPIE.6272E..77C}.
At that time only a long term
trend was appearing from RV data and our observations were aimed
at the direct detection of the companion responsible for the trend.
The acquisition and data reduction procedures are the same as
done for HD132563 in \citet{2011A&A...533A..90D}, with the differences that
all the images on HD106515 were taken using the broad band K' filter and
at the same rotation angle. 194 images were acquired and used in our analysis.
The projected separation between HD106515 A and B results of 6.897$\pm$0.015 arcsec and the
position angle of 267.07$\pm$0.12 deg.
Detection limits in K band magnitude difference were transformed in mass
limits using the mass-luminosity relation by \citet{2000A&A...364..217D}
at $M_{K}<9.5$ and the 5 Gyr theoretical isochrone by \citet{2000ApJ...542..464C}
at fainter magnitudes.

\section{Stellar properties}
\label{s:star}

\object{HD 106515} (HIP 59743, GJ 9398, ADS 8477) is a pair formed 
by two similar stars
slightly less massive than the Sun.
The main properties of the components of HD 106515 are summarized in 
Table~\ref{t:star_param}.

\begin{table}[h]
   \caption[]{Stellar properties of the components of HD 106515.}
     \label{t:star_param}

       \begin{tabular}{lccc}
         \hline
         \noalign{\smallskip}
         Parameter   &  \object{HD~106515~A} & \object{HD~106515~B}  & Ref. \\
         \noalign{\smallskip}
         \hline
         \noalign{\smallskip}

$\alpha$ (2000)          &  12 15 06.567 &  12 15 06.103&  1   \\
$\delta$ (2000)          & -07 15 26.38 &   -07 15 26.61 &  1   \\
$\mu_{\alpha}$ (mas/yr)  & \multicolumn{2}{c}{-249.67 $\pm$ 0.91}  & 2   \\
$\mu_{\delta}$ (mas/yr)  & \multicolumn{2}{c}{-52.29 $\pm$ 0.74}  & 2   \\
RV     (km/s)            &    20.66$\pm$ 0.11 &    19.94$\pm$0.11 & 3 \\
$\pi$  (mas)             & \multicolumn{2}{c}{28.42 $\pm$ 0.96}   & 2   \\
$d$    (pc)              & \multicolumn{2}{c}{ 35.2$\pm$1.1  }  & 2 \\
$U$   (km/s)             & \multicolumn{2}{c}{  -28.0  }  & 4 \\
$V$   (km/s)             & \multicolumn{2}{c}{  -38.7  }  & 4 \\
$W$   (km/s)             & \multicolumn{2}{c}{    4.6  }  & 4 \\
 & &  &   \\
V                        &  7.960$\pm$0.005  &   8.234$\pm$0.007  & 5 \\
$\Delta V$               & \multicolumn{2}{c}{  0.272$\pm$0.003  }  & 5 \\
B-V                      & \multicolumn{2}{c}{  0.815$\pm$0.003  }  &  1\\
B-V                      &      0.793$\pm$0.021    &    0.830$\pm$0.034   & 5 \\
V-I                      &      \multicolumn{2}{c}{  0.83$\pm$0.02  }    & 1 \\
$H_{p}$ scatter          & \multicolumn{2}{c}{0.011$^{\mathrm{a}}$} & 1 \\
$J_{2Mass}$              &  6.585$\pm$0.024       &  6.746$\pm$0.030  & 6 \\
$H_{2Mass}$              &  6.218$\pm$0.046       &  6.362$\pm$0.034  & 6 \\
$K_{2Mass}$              &  6.151$\pm$0.026       &  6.267$\pm$0.017  & 6 \\
NUV magnitude            &     \multicolumn{2}{c}{14.02}  & 7 \\
FUV magnitude            &     \multicolumn{2}{c}{--}  & 7 \\
 & & &  \\
$T_{eff}$ (K)            &  5232    &  5073    &  5 \\ %FEROS
$\Delta T_{eff}(A-B)$ (K)& \multicolumn{2}{c}{   164$\pm$21  } &  5 \\ %FEROS
$\log g$                 &  4.31      &   4.32      &  5 \\ %FEROS
 & & &    \\
$\log R^{'}_{HK}$        &  -5.04         &  -5.07       &  3 \\ %FEROS
$\log R^{'}_{HK}$        &  -5.07         &              &  8 \\ %Gray
$\log R^{'}_{HK}$        &  -5.06         &  -5.04       &  9 \\ %Arriagada

$ v \sin i $ (km/s)      &   0.6          &   1.3        &  3 \\  %FEROS
 & & &   \\
${\rm [Fe/H]}$           &   0.01  &   0.00  &  5 \\ %FEROS
$\Delta {\rm [Fe/H]}(A-B)$& \multicolumn{2}{c}{ 0.009$\pm$0.017  } &  5 \\ %FEROS
 & & &   \\
${\rm Mass} (M_{\odot})$ &    0.91$\pm$0.03$^{\mathrm{b}}$  &  0.88$\pm$0.03$^{\mathrm{b}}$  &  4 \\
Age  (Gyr)               &\multicolumn{2}{c}{$\sim 4-8$}  &  4  \\

         \noalign{\smallskip}
         \hline
      \end{tabular}

References: 1 Hipparcos \citep{1997ESASP1200.....P}; 
            2 \cite{2007A&A...474..653V};
            3 \cite{2006A&A...454..553D};
            4 This Paper;
            5 \cite{2006A&A...454..581D};
            6 2MASS \citep{2006AJ....131.1163S};
            7 Galex \citep{2005ApJ...619L...1M};
            8 \cite{2003AJ....126.2048G};
            9 \cite{2011ApJ...734...70A}

\begin{list}{}{}
\item[$^{\mathrm{a}}$] A+B
\item[$^{\mathrm{b}}$] Derived using the PARAM web interface \citep{2006A&A...458..609D}
\end{list}
\end{table}

The metallicity of the components was studied by us using both SARG and FEROS 
spectra \citep{2004A&A...420..683D,2006A&A...454..581D}. 
The metallicity resulted close to solar.
The line-by-line abundance differential analysis did not reveal
significant abundance difference between the components.

The slow projected rotational velocity, low level of chromospheric activity
\citep{2006A&A...454..581D,2003AJ....126.2048G,2011ApJ...734...70A},
and lack of detection in ROSAT All Sky Survey \citep{2000IAUC.7432R...1V}
are all consistent with a
rather old age, at least as old as the Sun. Isochrone fitting does not allow 
to put firm constraints
on stellar age, because of long timescales of the evolution of stars less massive
than the Sun. The thin disk kinematic places an upper limit of about 8 Gyr.
The system age is then in the range between 4 to 8 Gyr.

\section{A massive planet around HD106515}
\label{s:hd1106515a}

\begin{figure}
   \includegraphics[width=9cm]{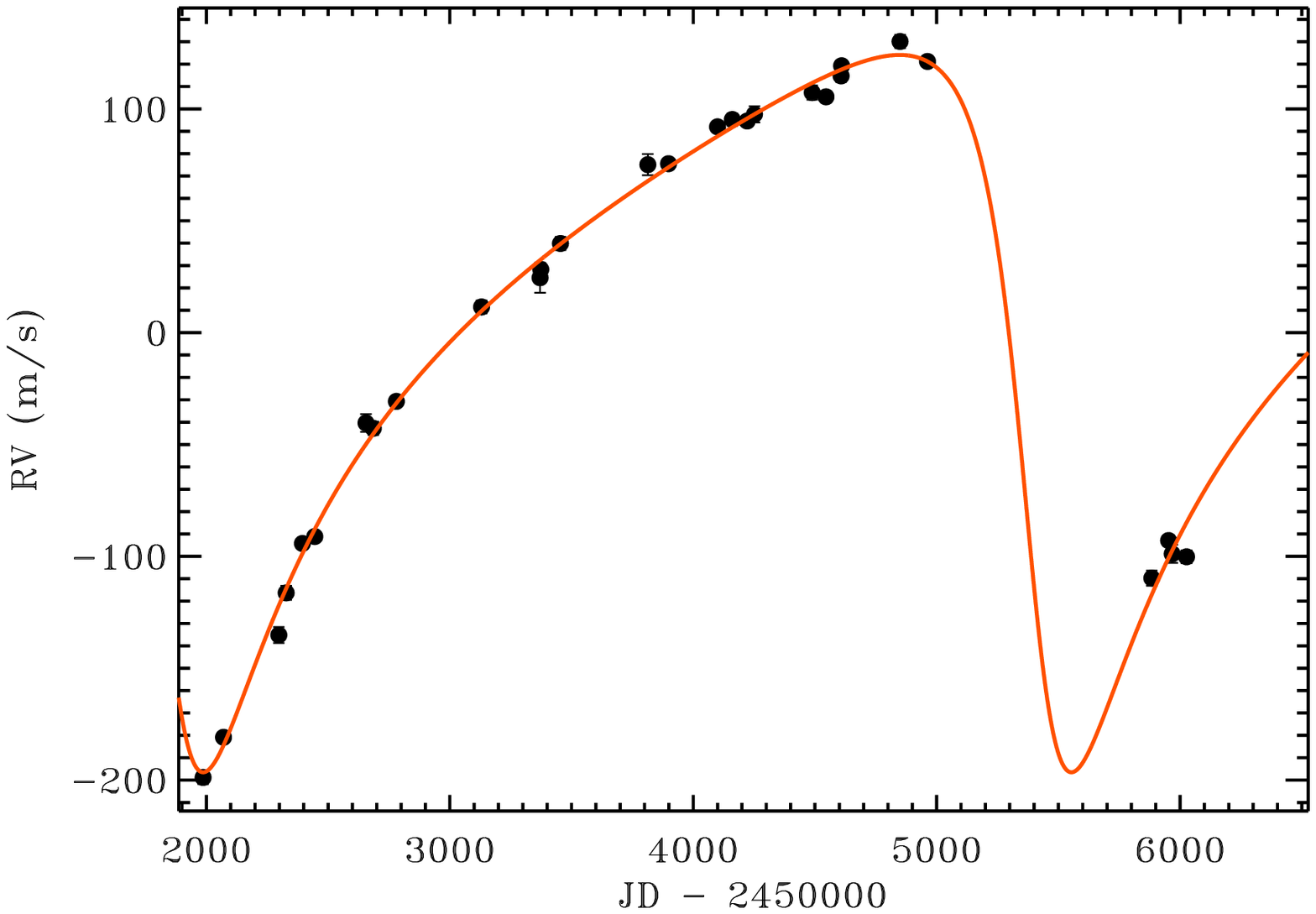}
   \includegraphics[width=9cm]{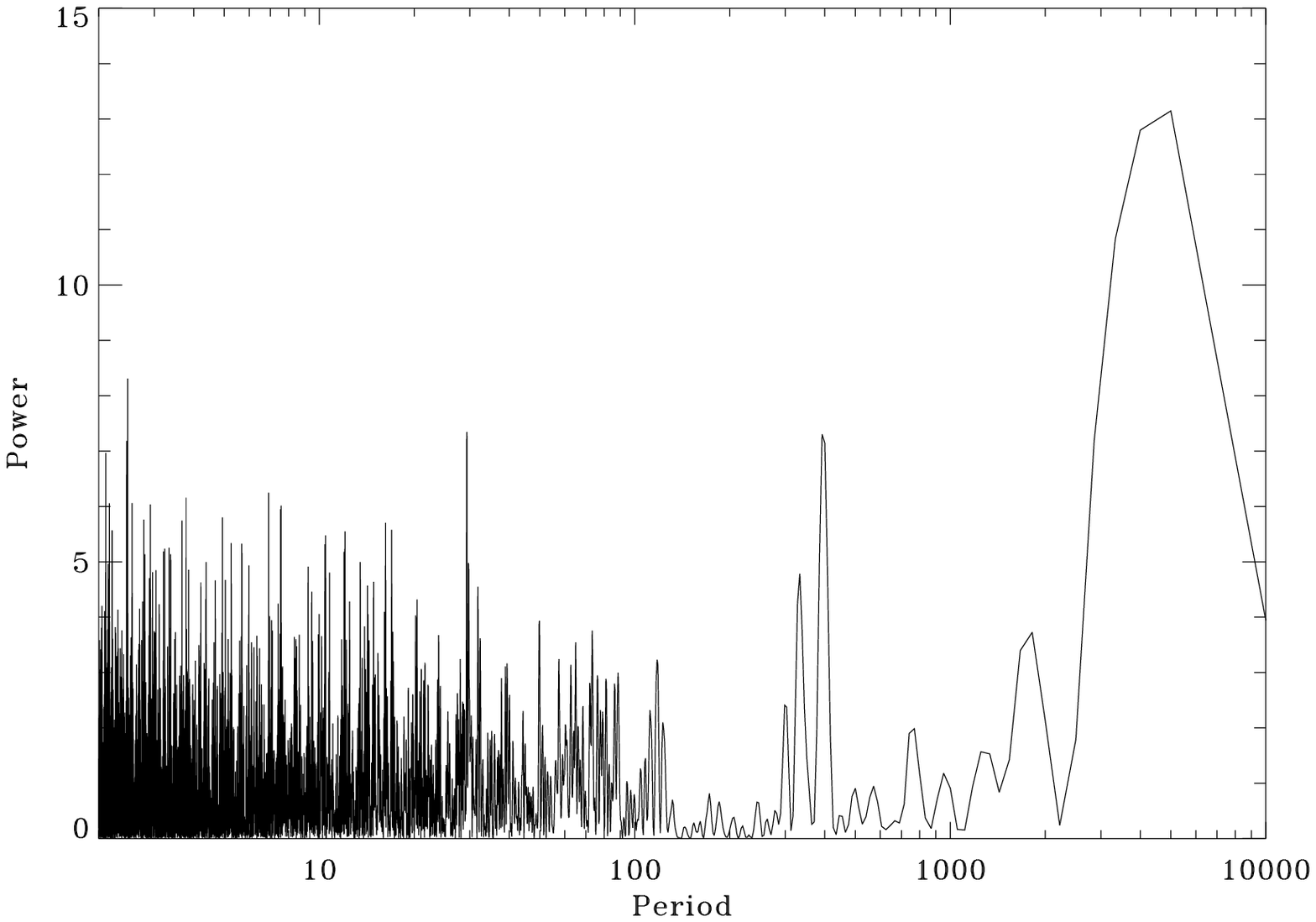}
      \caption{Upper panel: radial velocities of \object{HD 106515A} Overplotted
               the Keplerian best fit. 
               Lower panel: Lomb-Scargle periodogram of 
               the radial velocities.} 
         \label{f:rv_hd106515a}
   \end{figure}

\begin{figure}
   \includegraphics[width=9cm]{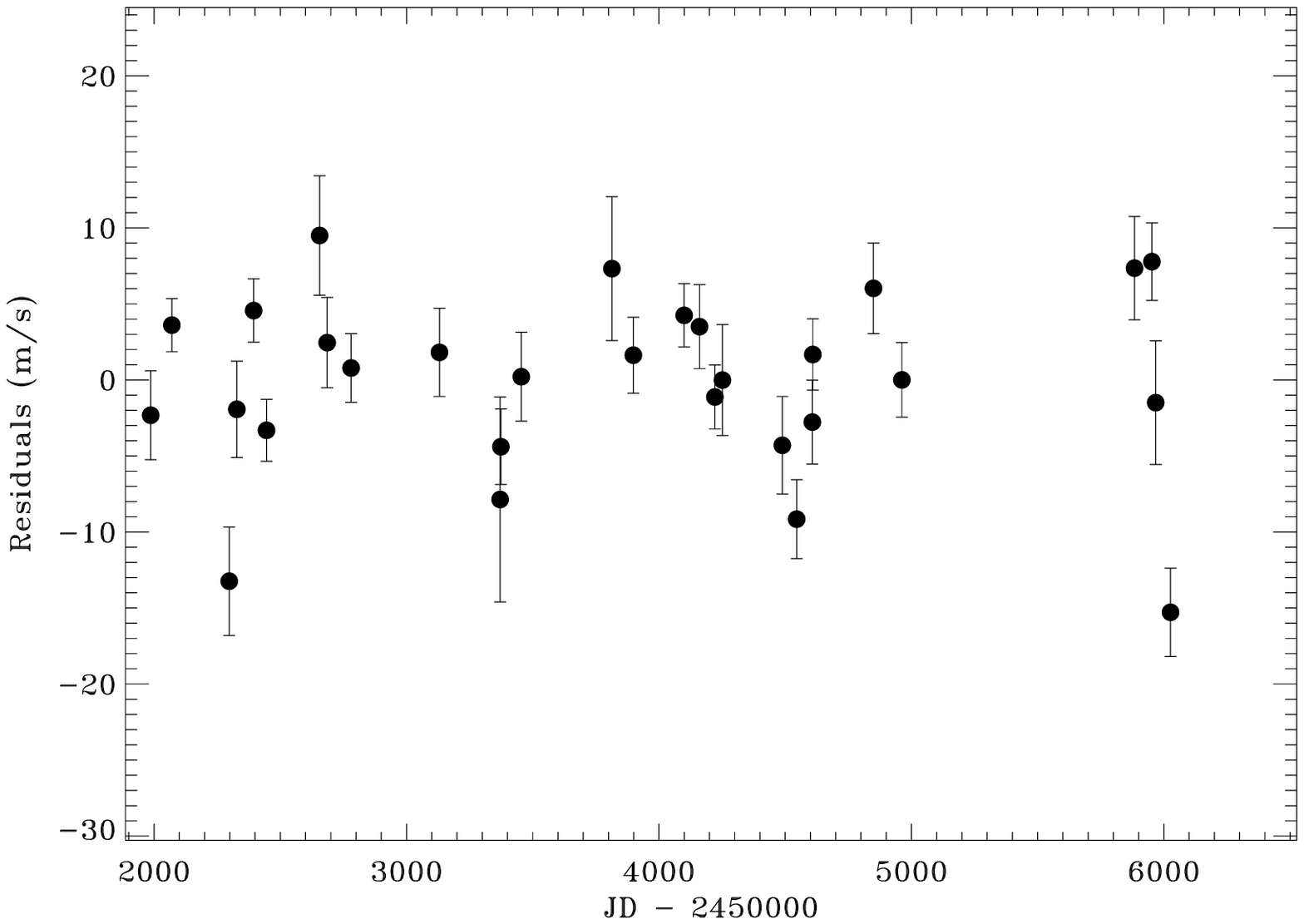}
   \includegraphics[width=9cm]{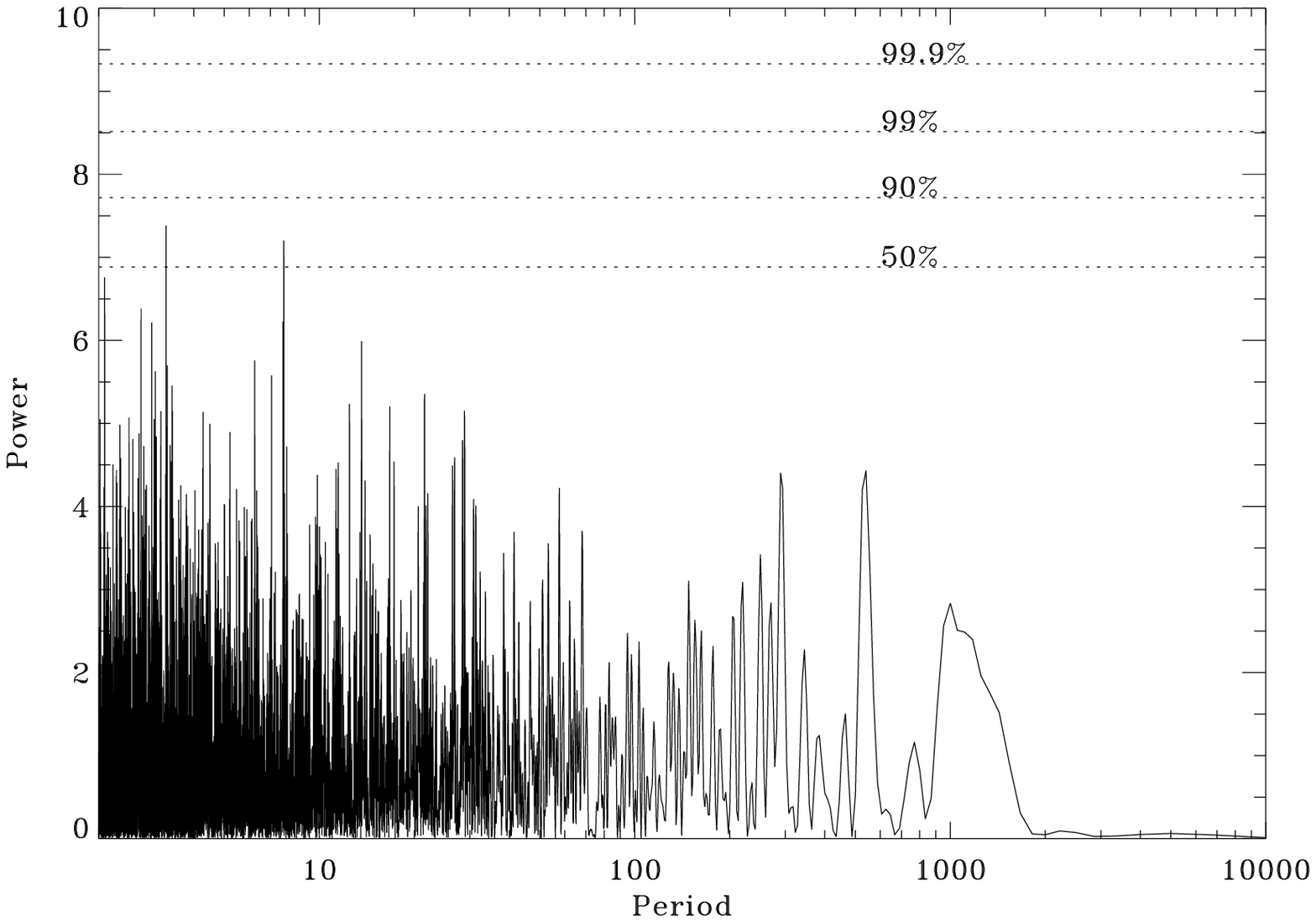}
      \caption{Upper panel: Residuals from best fit orbit vs time
               Lower panel: Lomb-Scargle periodogram of residuals from the Keplerian best fit of the
               radial velocities of \object{HD106515A}, with false alarm probability levels from
               bootstrap simulation overplotted.}        
         \label{f:res}
   \end{figure}

The RV time-serie of HD~106515A shows a long period modulation clearly detectable by eye
(Fig.~\ref{f:rv_hd106515a}). 
Lomb-Scargle periodogram confirms the long periodicity and the lack of shorter
periodicities in the data.
The low level of chromospheric activity guarantees the Keplerian
origin of the RV variations observed for HD~106515A.
An expected activity jitter of 3.5 m/s was derived using
\citet{2005PASP..117..657W} calibration.
We then performed Keplerian fitting using a Levenberg-Marquardt least-squares fit
algorithm as in \cite{2011A&A...533A..90D}.
The parameters are similar to those derived by \cite{2011arXiv1109.2497M}. Both
orbital parameters are listed in Table \ref{t:fit}. 
The residuals from the best fit Keplerian orbit
have an rms of about 6 m/s, and the periodogram does not show indication
of additional significant periodicities (Fig.~\ref{f:res}).
We then conclude that HD~106515A is orbited by a companion whose minimum mass is 
in the planetary regime. This should then be identified as \object{HD~106515Ab}.
In the following section we will exploit the available astrometric data to 
further constrain the mass of the companion.

\begin{table}
   \caption[]{Orbital parameters and results of fitting for RV of HD106515A.}
     \label{t:fit}
       \centering

 \begin{tabular}{lcc}
         \hline
         \noalign{\smallskip}
         Parameter  &  Our fit & Mayor et al.~2011 \\      
         \noalign{\smallskip}
         \hline
         \noalign{\smallskip}
Period (d)      &  3567$\pm$14     & 3630  \\
K (m/s)         &  160$\pm$3       & 174   \\
e               &  0.57$\pm$0.01   & 0.60  \\
$\omega$ (deg)  &  124$\pm$14      &  --   \\
T0 (JD-2450000) &  1844$\pm$27     &  --   \\
msini ($M_{J}$) &  9.33$\pm$0.16   & 10.50 \\
a (AU)          &  4.43$\pm$0.01   &  --   \\
rms res  (m/s) &    6.0           &  --   \\
      \noalign{\smallskip}
         \hline
      \end{tabular}

\end{table}

\begin{figure}
   \includegraphics[width=9cm]{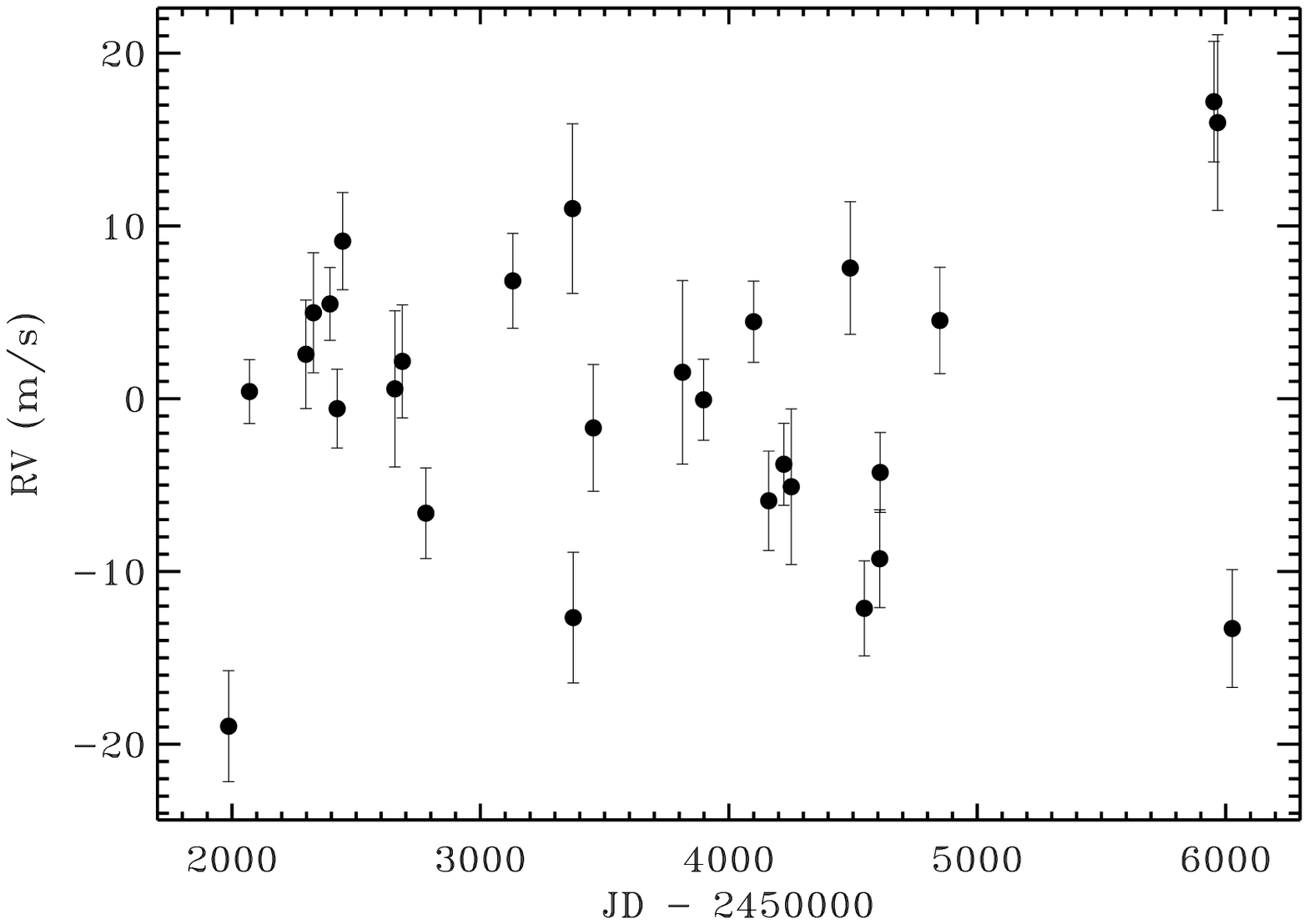}
   \includegraphics[width=9cm]{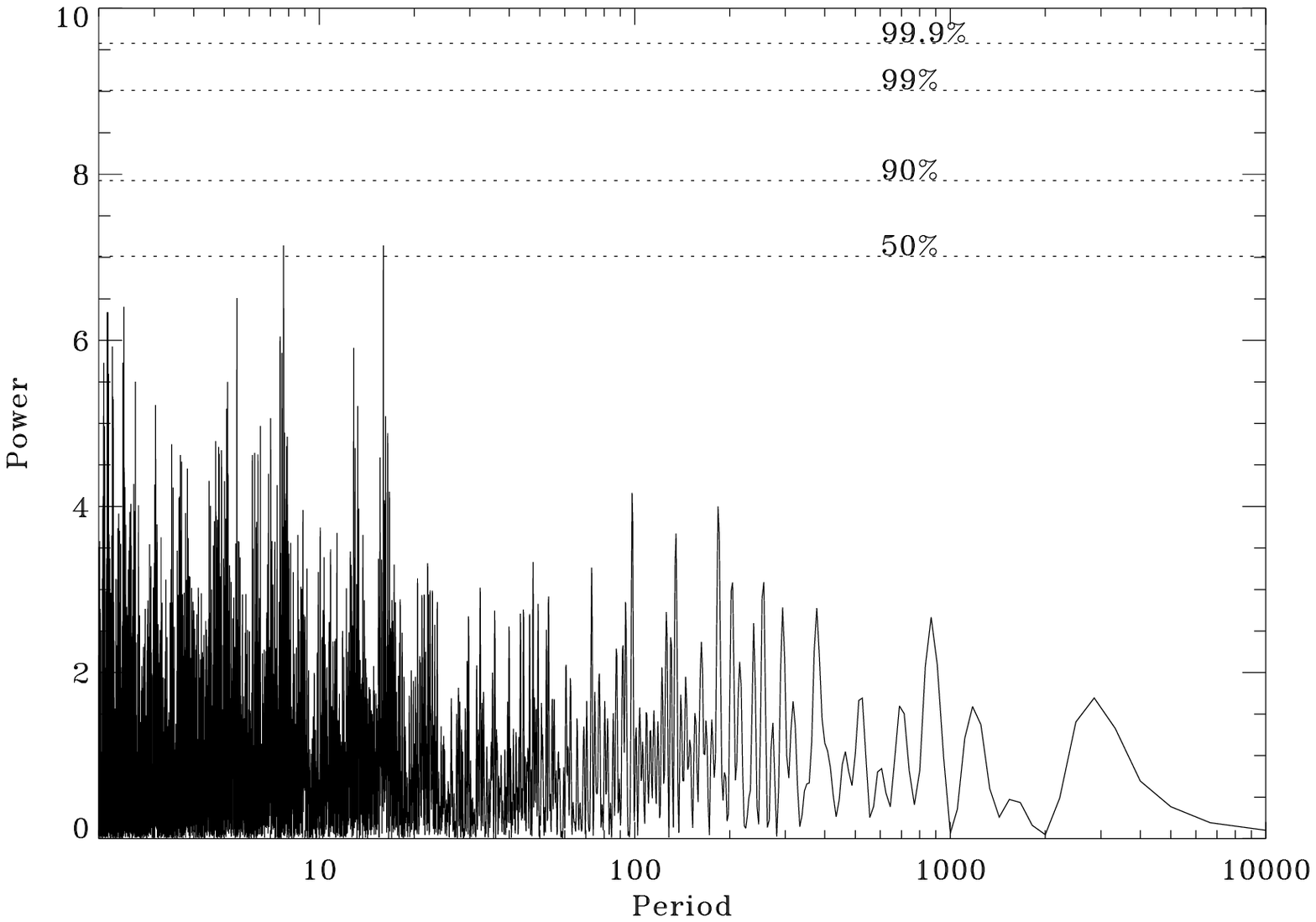}
      \caption{Upper panel: RV time serie of HD106515B
               Lower panel: Lomb-Scargle periodogram of RVs and false alarm probability levels from
               bootstrap simulation.}        
         \label{f:rvb}
   \end{figure}

The RV curve of the wide companion HD~106515B shows instead a scatter only 
slightly larger than internal errors (rms 8.7 m/s), without significant periodicities
or long term trends (Fig.~\ref{f:rvb}).

\section{Search for astrometric signature}
\label{s:astrometry}

Considering the parameters from the RV orbit and the distance to the system,
the amplitude of the astrometric signature for the RV minimum mass
is about 1.2 mas.
While we do not expect to be able to detect such astrometric motion in current data,
it is possible to have a significant detection in case the orbit is seen close to pole-on and
the actual mass is significantly larger than the projected mass.

\begin{figure}
   \includegraphics[width=9cm]{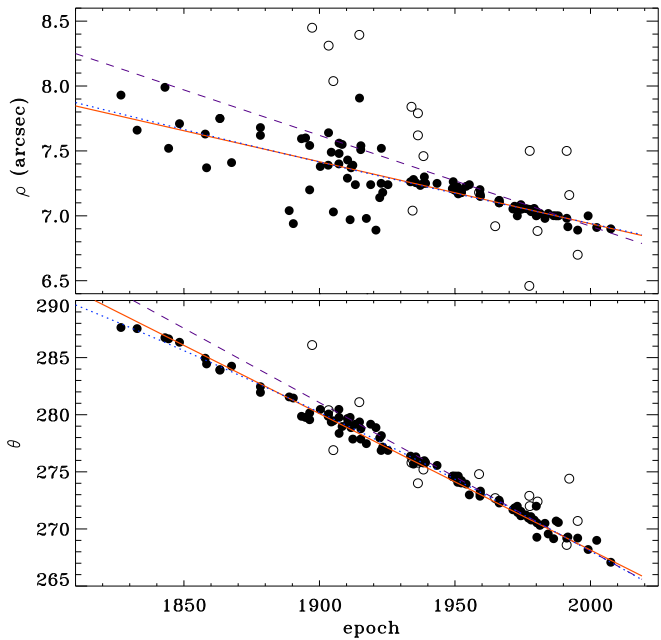}
      \caption{Upper panel: projected separation vs time for HD106515. Bottom panel:
               position angle vs time. In both panels filled circles are the measurements
               kept in the fit and the empty circles are those removed as being outliers in either
               projected separation or position angle.
               In both panels, red continuous line is the linear fit imposing
               passage through Hipparcos measurement, blue dotted line is the quadratic
               fit, purple dashed line is the slope as measured from Hipparcos (baseline 3.25 yr; epoch
               1991.25)
               extrapolated to the whole baseline of available data.}        
         \label{f:rhotheta}
   \end{figure}

We considered all the relative astrometry measurements available in Washington Double
Star catalog (kindly provided by B. Mason), that span about 180 yr.
Long period trends are clearly seen in both position angle and projected separation,
as expected from the orbital motion of the wide pair (Fig.~\ref{f:rhotheta}).
The observed slopes will be used in Sect.~\ref{s:binorbit} to constrain the binary orbit.
Short term slopes as measured with Hipparcos are consistent with those based on the full dataset,
supporting the lack of additional high amplitude astrometric perturbations on the timescales
comparable to the mission lifetime.

\begin{figure}
   \includegraphics[width=9cm]{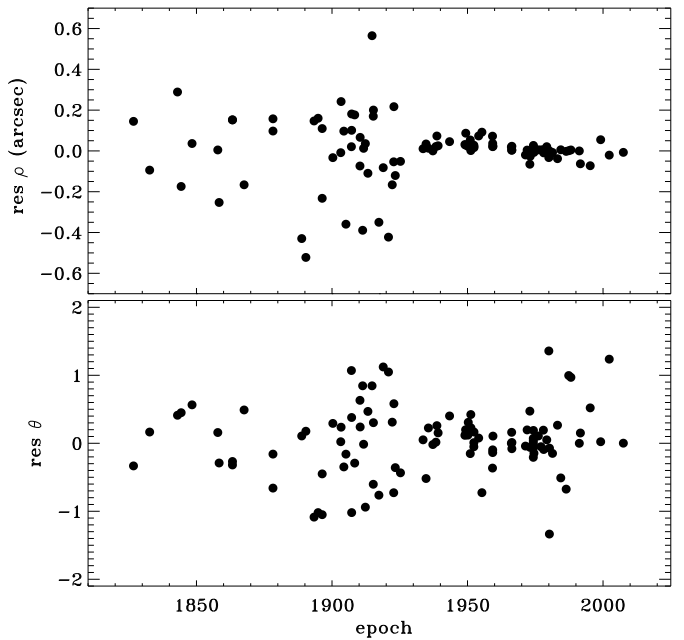}
      \caption{Residuals from quadratic fit in Fig.~\ref{f:rhotheta}. Upper panel: projected separation.
               lower panel: position angle.}        
         \label{f:resrhotheta}
   \end{figure}

Residuals from long term quadratic slope show a much larger scatter before 1930 (Fig.~\ref{f:resrhotheta}). 
We therefore considered only data taken after 1930 for the search of astrometric signatures due to the RV
companion.
Lomb Scargle periodograms of residuals in X and Y coordinates have very low power at period close
to that of RV orbit.
A possible periodicity is revealed at about 70 yr, especially in Y coordinate (the modulation
can also be seen by eye in the projected separation vs time plot).
Given the dense sampling of the astrometric measurements, this periodicity should not be related
to the RV companion.

\begin{figure}
   \includegraphics[width=9cm]{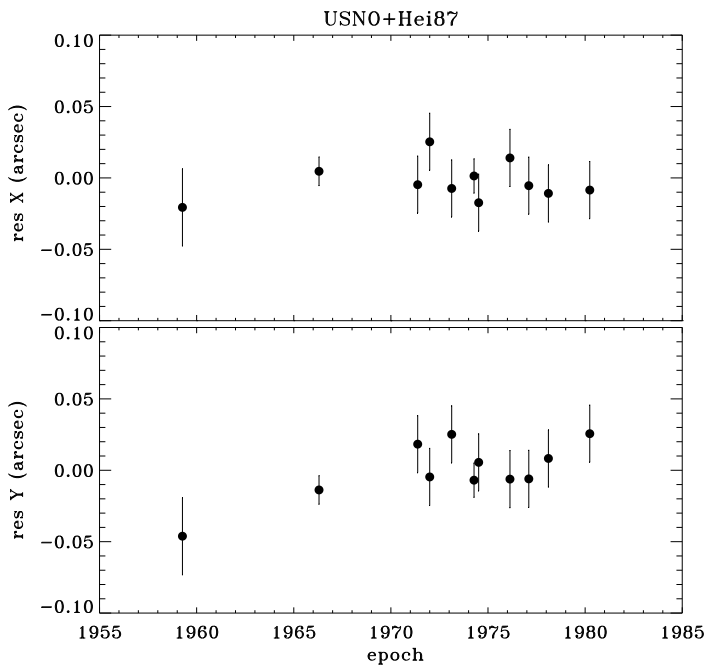}
      \caption{Residuals from full-data quadratic fit in X and Y coordinates. Only
                data from USNO (yearly averages) and Heintz 1987 are shown.}        
         \label{f:resusno.ps}
   \end{figure}

If we limit our analysis to data from USNO 
\citep{1963PUSNO..18....1J,1969PUSNO..18.....K,1978PUSNO..24....7J}
and \cite{1987ApJS...65..161H} (Fig.~\ref{f:resusno.ps}), to have 
a more homogeneous dataset
and reduce the impact of absolute calibration errors that might be present when using different
instrumentation, we have that the residuals have a dispersion of 13 mas in X and 20 mas in Y,
and of 14 and 15 mas in X and y respectively when taking the 70yr periodicity mentioned above into account.
These data cover 22 years. 
We then estimate an upper limit on the astrometric amplitude with 10 yr period of about
30 mas, comparing the expected astrometric motion of the companion for
different masses with the USNO and \cite{1987ApJS...65..161H} dataset. 
This corresponds to a limit in mass of about $0.25~M_{\odot}$.

\begin{figure}
   \includegraphics[width=9cm]{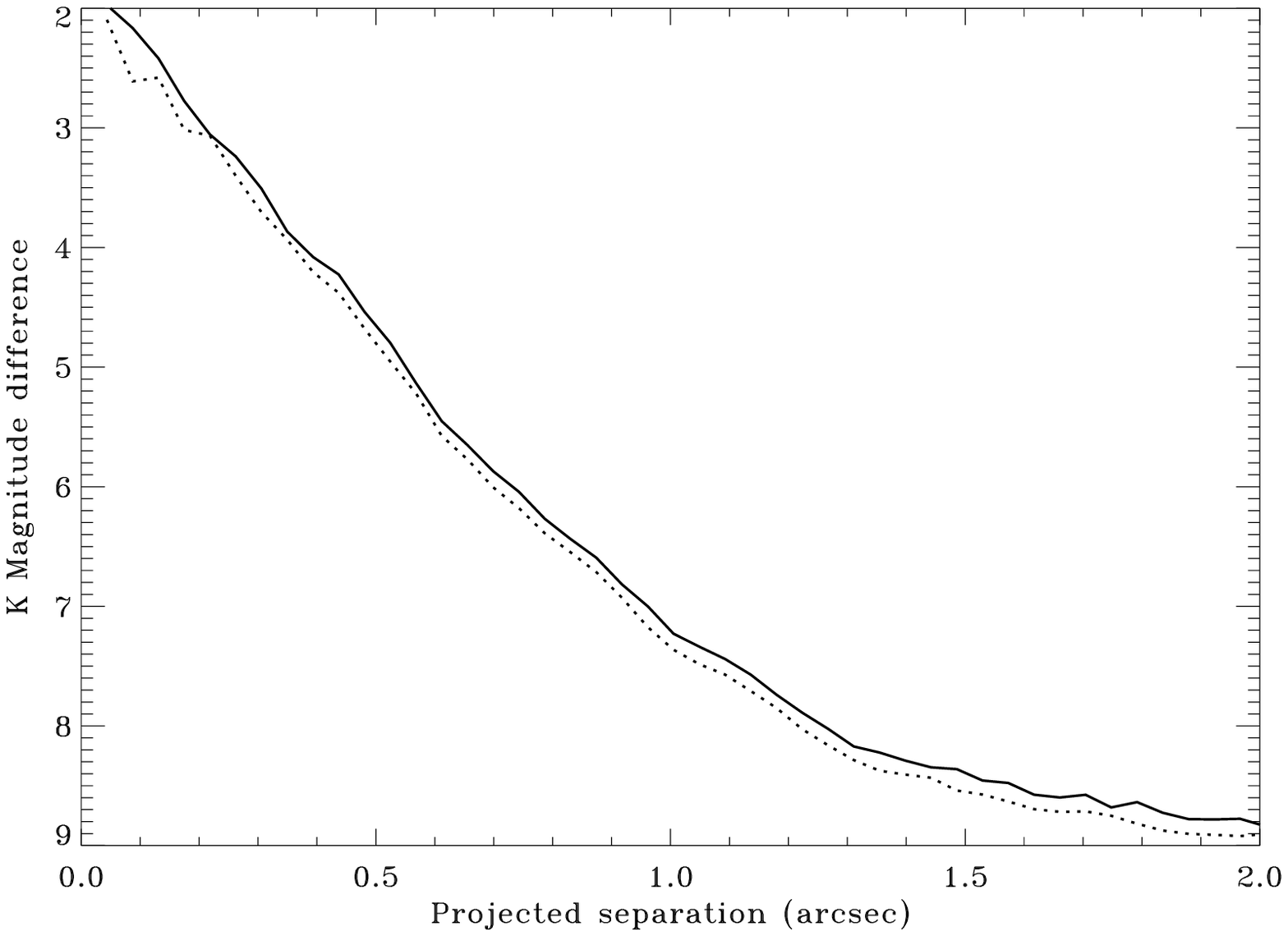}
   \includegraphics[width=9cm]{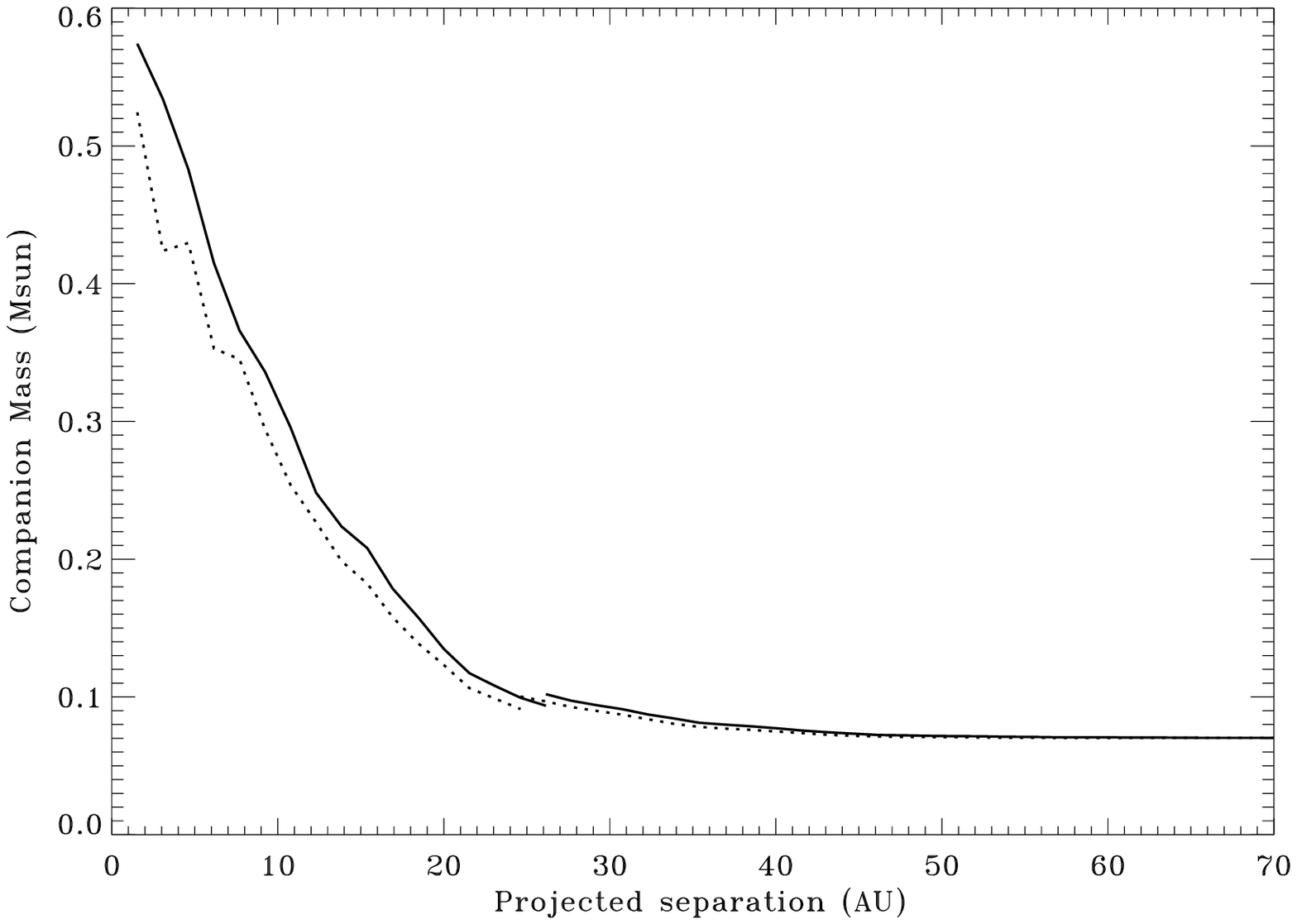}
      \caption{Detection limits for stellar companions around the components of HD106515
             from AdOpt@TNG images. Upper panel: $\Delta K$ vs projected separation in arcsec;
             Lower panel: Companion mass vs projected separation in AU. In both plots continuous line
             represent detection limits for HD106515A and dotted lines those for HD106515B}        
         \label{f:adoptlimits}
   \end{figure}

The tentative 70 yr periodicity with its possible amplitude of about 50 mas would
correspond to a very low mass star of about $0.1~M_{\odot}$.
The corresponding semimajor axis would be of about 17 AU, i.e. 0.5 arcsec on the sky.
Our images taken with AdOpt@TNG do not reveal stellar companions with masses
larger than $0.15-0.2~M_{\odot}$ at such a projected separation either around HD106515A or
around HD106551B (Fig.~\ref{f:adoptlimits}), therefore, the non-detection is not conclusive.
The RV semiamplitude of a $0.1~M_{\odot}$  companion in 70yr orbit in edge-on 
circular orbit is about 700 m/s, with RV difference that might reach
600 m/s over 10 years.
The real slope might be significantly smaller than this value, depending
on the actual phase of the orbit at the time of the observations, 
inclination, eccentricity. 
The data of HD106515B does not reveal significant long term trends  
(Fig.~\ref{f:rvb}), making very unlikely
that such an additional companion is orbiting around this star.
The RV data for HD106515A are less adequate for the study of
additional long term trends as the temporal baseline slightly exceeds 
one orbital period of HD106515Ab and the sampling after 2009 is poor.
Nevertheless, we do have specific indications of the presence of an additional
long term trend in the data.

 \begin{figure}
   \includegraphics[width=9cm]{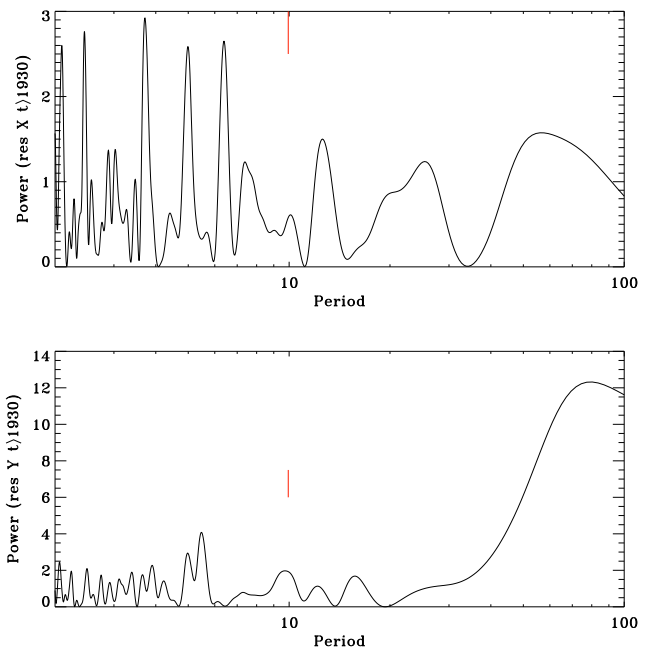}
      \caption{Lomb-Scargle periodogram of the residuals from full-data quadratic fit in X and Y coordinates. 
                Only data after 1930 were considered because of their higher precision. The short vertical lines at 10yr mark
                the RV period.}        
         \label{f:scargleresxy.ps}
   \end{figure}

We also note that no measurable differences are found in the stellar proper
motion between the new reduction of the short-term Hipparcos astrometric
data for HD 106515
\citep{2007A&A...474..653V} and the long-term Tycho-2 data \citep{2000A&A...355L..27H}.
The proper motion values in the two catalogs agree well with each other
within the quoted
uncertainties, thus no useful constraint can be obtained on the orbit
and mass ratio based on the $\Delta\mu$ technique \citep[e.g.][]{2005AJ....129.2420M}.

\section{Binary orbit}
\label{s:binorbit}

To constrain the orbit of the HD106515 system, we considered the long term 
relative astrometry (Sect.~\ref{s:astrometry}) and the RV difference between the 
components.
To derive this latter quantity, we measured the radial velocities of both components using
the stellar template of HD106515A.
Taking the orbital motion of the massive planet around the primary into account, 
such a difference results of $\Delta RV_{A-B} = 739$ m/s.
Internal errors derived from the error of the mean of the RV of the individual components
and number of spectra and including uncertainty in the planet orbit are within 10 m/s.
Systematic effects due to spectral mismatch were estimated by \cite{2002ApJS..141..503N} to 
be of the order of 100 m/s
when using solar spectrum as template for the derivation of the absolute RVs of FGK stars.
This effect would be significantly smaller in our case thanks to the small temperature difference
between the components and the very similar metallicities and projected rotational velocities. 
The difference in the convetive blueshift and gravitational redshift of the components amount to
just 24 m/s following Eq. 3 of \cite{2002ApJS..141..503N}. Overall, the true error of our determination
is likely within 50 m/s.

\begin{figure}
   \includegraphics[width=9cm]{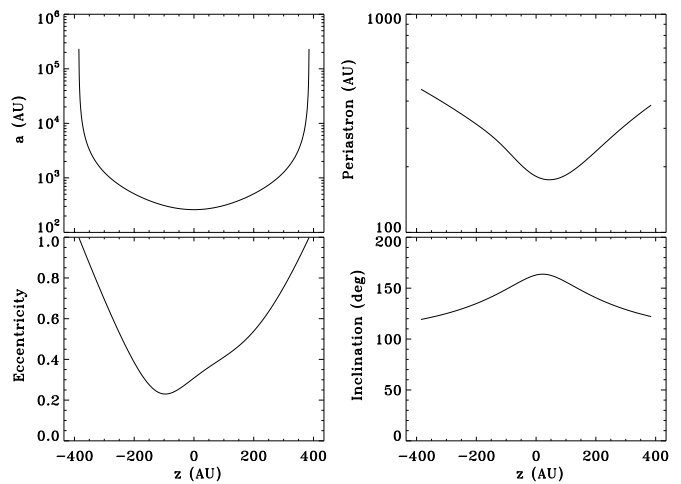}
      \caption{Possible orbital parameters of the HD106515 system for various separations
               along the line of sight at present epoch. Top-left: semimajox axis; top-right: periastron of
               the orbit; bottom-left: eccentricity; bottom-right: inclination.}        
         \label{f:binorbit}
   \end{figure}

Having the relative position and velocities on the plane of the sky and the 
relative velocity along the line of sight, we then derived the family of possible bound solutions 
as a function of the separation between the components along the line of sight ($z$)
using the approach by \cite{1999PASP..111..321H} as in \cite{2011A&A...533A..90D}.
The results are displayed in Fig.~\ref{f:binorbit}.

\begin{figure}
   \includegraphics[width=9cm]{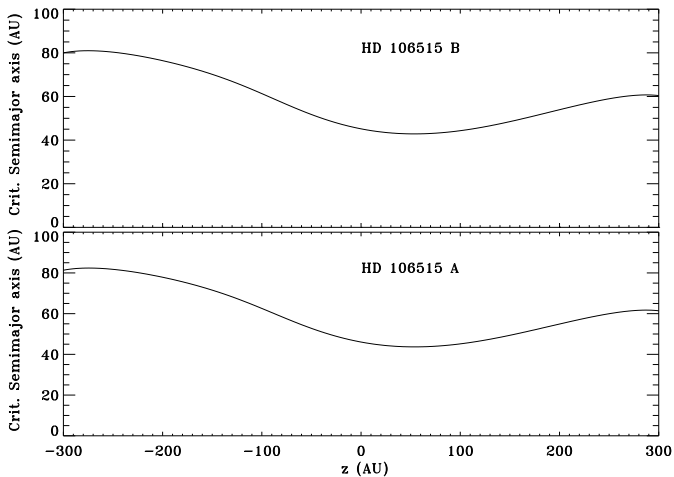}
      \caption{Critical semimajor axis for dynamical stability for planets around the components of HD106515,
               for various separations along the line of sight at present epoch and corresponding binary orbit
               parameters.}        
         \label{f:acrit}
   \end{figure}

Critical semimajor axis for dynamical stability of planets around each component, calculated following
\citet{1999AJ....117..621H}, are shown in Fig.~\ref{f:acrit} for the family of possible orbits
shown in Fig.~\ref{f:binorbit}.
This quantity is larger than 40 AU for all the orbits\footnote{At values of $|z|$ larger than 300 
the large eccentricity of the binary orbit makes the \citet{1999AJ....117..621H} equations no 
longer valid.}.
Therefore, the planet candidate is well within the stability boundaries for all possible orbits 
of the wide binary. 
The potential astrometric candidate discussed in Sect.~\ref{s:astrometry} is also well within the
stability zone. However, if real, its presence should alter the RV difference, the astrometric
trends and the mass of one component, with some effects on the family of possible binary orbits.

\section{Discussion}
\label{s:discussion}

The planet candidate around HD106515A, with its minimum mass of about $9.5~M_{J}$, 
is one of the few with projected masses close to deuterium burning limit.
From available data, we did not detect any additional planets in the system.
The high eccentricity of its orbit and the solar-like metallicity of its parent star
are in agreement with the differences in the statistical properties of planets
below and above 4 $~M_{J}$ found by \cite{2007A&A...464..779R}.

The planet host has a stellar companion of similar mass, then HD106515Ab adds to the growing
census of exoplanets in multiple systems \citep{2007A&A...462..345D,2012A&A...542A..92R}.
The binary separation is quite wide and plausible orbits leave dynamically stable zones
up to 40-80 AU around the stars. This suggests a limited impact of the companion on the planet
properties but the moderate eccentricity might also be linked to Kozai interactions,
which are effective even for widely separated companion considering the old age of the system
\citep{2005ApJ...627.1001T}. We also found from the analysis of the relative
astrometry tentative indication of an additional object with a period of 70 yr.
From our data, we can not confirm the reality of this object, that might be
a very low mass star, and infer the component around which it should be orbiting.

The presence of a well separated companion with similar properties allowed us to perform
a sensitive differential abundance analysis \citep{2004A&A...420..683D,2006A&A...454..581D}.
The lack of significant metallicity differences between the components extends
the previous finding that large alterations of chemical abundances somewhat linked to
the presence of planets are not a common event \citep{2006A&A...454..581D,2011A&A...533A..90D}.
There are 8 binary systems with giant planets suitable for the comparison of chemical
abundances \citep[HD 106515 and the 7 listed 
in Table 8 of][]{2011A&A...533A..90D}. All of them have $\Delta$ [Fe/H] $<0.05$ dex, which is
significantly smaller than the typical difference in metallicity between giant planet hosts 
and nearby field stars \citep[$\Delta \rm{[Fe/H]} \sim 0.25$ dex, see e.g.][]{2005ApJ...622.1102F,2004A&A...418..989N}. 
This supports the primordial origin for the metallicity enhancement of stars with giant planets
\citep{2005ApJ...622.1102F}.

The binarity of HD106515 represents also a suitable opportunity for the true mass determination
HD106515Ab, thanks to the reference provided by HD106515B.
The expected astrometric amplitude is of about 1.2 mas for the minimum mass
and about 10 mas at stellar/substellar boundary.

From available relative astrometry, the orbital motion of the wide pair is clearly detected.
From the analysis of residuals from the long term trend at epochs 1959-1980 where several high-quality
data are available we put an upper limit to the mass of the companion of about
$0.25~M_{\odot}$.
Much better astrometric precision can be obtained by more recent instrumentation.
HD106515 is an ideal target for differential astrometry using AO 
systems \citep{2009MNRAS.400..406H,2010EAS....42..179R}
and new interferometric instruments like PRIMA \citep{2011EPJWC..1607005Q}.
Subsequently, the combination of ground-based radial velocities and Gaia
high-precision space-borne astrometric data might prove decisive 
\citep[e.g.,][and references therein]{2011EAS....45..273S}.

Therefore, there are very promising perspectives for a true mass determination of
the companion of HD106515A in the coming years, then removing the ambiguity due to projection
effects. The availability of true masses is relevant for a better understanding of the 
high-mass tail of the planetary mass function  and the transition between planets and brown dwarfs.
The direct detection of the companion is instead more challenging even for next generation
planet finders as SPHERE or GPI \citep{2010ASPC..430..231B}, because of the small projected 
separations ($<0.2$ arcsec) and faint luminosities implied by the old age of the system, unless
the orbit is seen nearly pole-on and its mass is significantly larger than the minimum mass.

\begin{acknowledgements}

This research has made use of the 
SIMBAD database, operated at CDS, Strasbourg, France. 
This research has made use of the Washington Double Star Catalog 
maintained at the U.S. Naval Observatory. 
We thank the TNG staff for contributing to the observations and the TNG TAC for
generous allocation of observing time.
We thank B. Mason for providing the astrometric data collected in the Washington Double Star Catalog.
This work was partially funded by PRIN-INAF 2008 and PRIN-INAF 2010.

\end{acknowledgements}

\bibliography{hd106515}
\bibliographystyle{aa}

\Online

\begin{table}
   \caption[]{Differential radial velocities of HD~106515~A}
     \label{t:rva}
       \centering
       \begin{tabular}{ccc}
         \hline
         \noalign{\smallskip}
         HJD -2450000  &  RV & error  \\
                      &  m/s & m/s   \\
         \noalign{\smallskip}
         \hline
         \noalign{\smallskip}
 1986.5429    &  -198.8    &     2.9   \\
 2070.3741    &  -180.9    &     1.7   \\
 2297.7099    &  -135.2    &     3.6   \\
 2327.7449    &  -116.3    &     3.2   \\
 2394.3984    &   -94.2    &     2.1   \\
 2445.4047    &   -91.1    &     2.0   \\
 2655.6473    &   -40.4    &     3.9   \\
 2685.7174    &   -42.8    &     3.0   \\
 2780.4680    &   -30.7    &     2.3   \\
 3130.4615    &    11.5    &     2.9   \\
 3370.7863    &    24.5    &     6.7   \\
 3373.7381    &    28.3    &     2.5   \\
 3454.5974    &    39.8    &     2.9   \\
 3813.5819    &    75.1    &     4.7   \\
 3898.4270    &    75.5    &     2.5   \\
 4099.8020    &    92.0    &     2.1   \\
 4160.6046    &    95.3    &     2.8   \\
 4221.5300    &    94.6    &     2.1   \\
 4251.4168    &    97.6    &     3.6   \\
 4488.7506    &   107.3    &     3.2   \\
 4545.4910    &   105.4    &     2.6   \\
 4607.4763    &   114.7    &     2.8   \\
 4609.4644    &   119.3    &     2.3   \\
 4849.6782    &   130.1    &     3.0   \\
 4962.4236    &   121.2    &     2.5   \\
 5883.7810    &  -109.7    &     3.4   \\
 5952.7662    &   -92.9    &     2.6   \\
 5967.6966    &   -98.9    &     4.1   \\
 6026.5722    &  -100.2    &     2.9   \\
         \noalign{\smallskip}
         \hline
      \end{tabular}

\end{table}

\begin{table}
   \caption[]{Differential radial velocities of HD~106515~B}
     \label{t:rvb}
       \centering
       \begin{tabular}{ccc}
         \hline
         \noalign{\smallskip}
         HJD -2450000  &  RV & error  \\
                      &  m/s & m/s   \\
         \noalign{\smallskip}
         \hline
         \noalign{\smallskip}
 1986.5544    &   -19.0    &     3.2   \\
 2070.3879    &     0.4    &     1.9   \\
 2297.7217    &     2.6    &     3.1   \\
 2327.7578    &     5.0    &     3.5   \\
 2394.5070    &     5.5    &     2.1   \\
 2423.3945    &    -0.6    &     2.3   \\
 2445.4169    &     9.1    &     2.8   \\
 2655.6592    &     0.6    &     4.5   \\
 2685.7307    &     2.2    &     3.3   \\
 2780.4806    &    -6.6    &     2.6   \\
 3130.4736    &     6.8    &     2.7   \\
 3370.7985    &    11.0    &     4.9   \\
 3373.7496    &   -12.7    &     3.8   \\
 3454.6096    &    -1.7    &     3.7   \\
 3813.5692    &     1.5    &     5.3   \\
 3898.4402    &    -0.1    &     2.3   \\
 4099.8161    &     4.5    &     2.3   \\
 4160.6168    &    -5.9    &     2.9   \\
 4221.5415    &    -3.8    &     2.4   \\
 4251.4282    &    -5.1    &     4.5   \\
 4488.7628    &     7.6    &     3.8   \\
 4545.5030    &   -12.1    &     2.8   \\
 4607.4885    &    -9.3    &     2.8   \\
 4609.4782    &    -4.3    &     2.3   \\
 4849.6896    &     4.5    &     3.1   \\
 5952.7777    &    17.2    &     3.5   \\
 5967.7085    &    16.0    &     5.1   \\
 6026.5845    &   -13.3    &     3.4   \\
         \noalign{\smallskip}
         \hline
      \end{tabular}

\end{table}

\end{document}